\documentclass[prd,superscriptaddress,nofootinbib,notitlepage]{revtex4-1}
\usepackage{epsfig,amsmath,amssymb,slashed}
\usepackage{graphicx}
\usepackage{color}
\usepackage{siunitx}
\usepackage{color}
\usepackage{bm}
\usepackage{subfig}
\newcommand{\bra}[1]{\langle #1|}
\newcommand{\ket}[1]{|#1\rangle}

\newcommand{\di}{{\rm d}}
\newcommand{\ii}{i}

\def\wT{{\widehat T}}
\def\wj{{\widehat j}}
\def\wJ{{\widehat J}}
\def\wK{{\widehat K}}
\def\wP{{\widehat P}}
\def\wA{{\widehat A}}
\def\wB{{\widehat B}}
\def\wQ{{\widehat Q}}

\def\wpsi{{\widehat{\psi}}}
\def\wrho{{\widehat{\rho}}}

\newcommand{\tr}{{\rm tr}}  
  
\newcommand{\e}{{\rm e}}

\newcommand{\p}{{\rm p}}
\newcommand{\x}{{\rm x}}
\newcommand{\y}{{\rm y}}

\newcommand{\wR}{\widehat{\mathcal{R}}}
\newcommand{\Deltaz}{\tilde\Delta}
\newcommand{\be}{\begin{equation}}
\newcommand{\ee}{\end{equation}}                                                                               
\newcommand{\bea}{\begin{eqnarray}}
\newcommand{\eea}{\end{eqnarray}}                                                                               
\begin{document}

\title{Quantum corrections to the stress-energy tensor in thermodynamic equilibrium
with acceleration} 

\author{F. Becattini}
\affiliation{Universit\`a di Firenze and INFN Sezione di Firenze, Florence, Italy}
\author{E. Grossi}
\affiliation{Universit\`a di Firenze and INFN Sezione di Firenze, Florence, Italy}

\begin{abstract}
We show that the stress-energy tensor has additional terms with respect to the ideal 
form in states of global thermodynamic equilibrium in flat spacetime with non-vanishing 
acceleration and vorticity. These corrections are of quantum origin and their leading 
terms are second order in the gradients of the thermodynamic fields. Their relevant 
coefficients can be expressed in terms of correlators of the stress-energy 
tensor operator and the generators of the Lorentz group. With respect to previous 
assessments, we find that there are more second order coefficients and that all 
thermodynamic functions including energy density receive acceleration and vorticity 
dependent corrections. Notably, also the relation between $\rho$ and $p$, that is 
the equation of state, is affected by acceleration and vorticity. We have calculated 
the corrections for a free real scalar field -- both massive and massless -- and 
we have found that they increase, particularly for a massive field, at very high acceleration 
and vorticity and very low temperature. Finally, these non-ideal terms depend on the 
explicit form of the stress-energy operator, implying that different stress-energy 
tensor of the scalar field -- canonical or improved -- are thermodynamically 
inequivalent.
\end{abstract}

\maketitle

\section{Introduction}
\label{intro}

It is common wisdom that the form of the relativistic stress-energy tensor in a 
thermodynamic equilibrium state has the ideal form:
$$
  T^{\mu\nu} = (\rho + p) u^\mu u^\nu - p g^{\mu\nu}
$$ 
where $\rho$ and $p$ are the energy density and pressure, thermodynamic functions of 
temperature $T$ and chemical potential $\mu$, and $u$ a constant four-velocity. 
In quantum statistical mechanics, the above expression corresponds to the renormalized
mean value \footnote{For free quantum fields, by renormalization we mean the use of normal 
ordering in the stress-energey tensor operator} of the quantum stress-energy tensor 
operator, built from local quantum fields, with the density operator:
\be\label{homo}
 \wrho = (1/Z) \exp [-\beta \cdot \wP + \zeta \wQ]
\ee
where $\beta = (1/T) u$ is a constant inverse temperature four-vector (or, simply, 
four-temperature), with $T = 1/\sqrt{\beta^2}$ being the proper (or comoving) temperature 
and $u$ the constant four-velocity, $\zeta=\mu/T$ is the ratio between proper chemical 
potential and proper temperature, $\wP$ the four-momentum operator and $\wQ$ an internal 
conserved charge:
\be\label{tideal}
  T^{\mu\nu}(x) = \tr ( \wrho \wT^{\mu\nu}(x))_{\rm ren} = \frac{1}{Z} \tr (\wT^{\mu\nu}(x) 
  \exp [-\beta \cdot \wP+ \zeta \wQ])_{\rm ren}  = (\rho + p) u^\mu u^\nu - p g^{\mu\nu} 
\ee
The form (\ref{tideal}) is dictated by the symmetries of the density operator (\ref{homo})
which is traslationally invariant and isotropic in the rest frame where $\beta = 
(1/T)(1,{\bf 0})$. 

However, the density operator (\ref{homo}), is not the only form 
of global thermodynamic equilibrium, which is, in general, a state where the entropy 
$S=-\tr (\wrho \log \wrho)$ is constant. For instance, it is well known \cite{landau,
vilenkin} that in non-relativistic quantum mechanics the operator:
\be\label{rotating}
  \wrho = (1/Z) \exp [-\widehat H /T_0 + \omega \widehat J_z/T_0]
\ee
where $T_0$ is a constant global temperature \footnote{The global temperature $T_0$ 
is a temperature measured by a thermometer at rest with the external observer.  
In general, it differs from the proper temperature $T$ measured by a comoving thermometer}, 
$\widehat H$ the hamiltonian and $\widehat J_z$ the angular momentum operator along 
some axis $z$, represents a globally equilibrated spinning fluid with angular velocity 
$\omega$. Similarly (see sect.~\ref{statmech}) the operator:
\be\label{accelerating}
  \wrho = (1/Z) \exp [-\widehat H /T_0 + a \widehat K_z/T_0]
\ee
$\widehat K_z$ being the generator of a Lorentz boost along the $z$ axis, represents 
a relativistic fluid with constant comoving acceleration along the $z$ direction 
and it is still an equilibrium distribution. These two cases belong to a more general 
class of thermodynamic equilibria which, in special relativity, are characterized by 
a four-temperature $\beta(x)$ field fulfilling the equation
\be\label{kill}
  \partial_\mu \beta_\nu + \partial_\nu \beta_\mu = 0  
\ee
which means that the four-temperature is a Killing vector field. 
We will show how for such thermodynamic equilibrium states, with the appropriate
treatment in quantum relativistic statistical mechanics, the ideal form of the 
stress-energy tensor gets quantum corrections -- vanishing in the $\hbar \rightarrow 0$ 
limit -- whose leading terms are proportional to the squared gradients of $\beta$, 
which in turn can be expressed in terms of the acceleration $a^\mu$ and the 
vorticity $\omega^\mu$ fields (see sect.~\ref{expansion} for definitions): 
$$
  T^{\mu\nu} = (\rho + p) u^\mu u^\nu - p g^{\mu\nu} + \hbar^2 \left( {\cal O}(a^2)+
  {\cal O}(\omega^2)+{\cal O}(a\omega) \right)
$$
As we will see, these corrections are normally tiny but they can become relevant 
under specific circumstances and, moreover, they are not microscopic in the sense 
of being relevant only at very small scales.   

The appearance of these terms is somehow in contrast to the widespread belief that 
deviations from the ideal form (\ref{tideal}) can only arise in presence of dissipative 
processes. In fact, the existence of such terms has been pointed out by a classification 
of second order gradient corrections of the stress-energy tensor in conformal hydrodynamics 
\cite{roma1,roma2} also by means of kinetic theory~\cite{denicol} and some coefficients,
denoted as thermodynamic in view of their survival at equilibrium, have been 
calculated in ref.~\cite{moore} for conformal field theories. 

In this paper, we show that the occurrence of non-dissipative corrections to the 
ideal form of the stress-energy tensor is a general fact which is related to the 
very notion of equilibrium in quantum relativistic statistical mechanics. Moreover,
these corrections result from the expansion of the density operator and their 
form is not assumed {\it a priori} like in the Landau-frame based gradient expansion.
At equilibrium, they have simple and suggestive expressions as correlators of the 
stress-energy tensor with the generators of the Lorentz group. The proper energy 
density expression is also modified, as well as the relation between energy density and 
pressure, that is the equation of state. It is an almost straigthforward consequence
that these corrections will extend to a curved spacetime. 

The paper is organized as follows: in section \ref{statmech} we obtain the form of 
the density operator of general thermodynamic equilibrium in quantum statistical mechanics
in flat spacetime. In section \ref{locality} we discuss the relation between local 
observables and the local value of the four-temperature field. In section \ref{expansion} 
we derive the form of the corrections to the ideal form of the stress-energy tensor 
as a perturbative expansion. In section \ref{freef} we calculate those quantum corrections 
in free scalar field theory. Finally, in sections \ref{discuss} and \ref{conclu} we 
discuss the most important physical consequences and draw the conclusions.

\subsection*{Notation}

In this paper we use the natural units, with $\hbar=c=K=1$.\\ 
The Minkowskian metric tensor is ${\rm diag}(1,-1,-1,-1)$; for the Levi-Civita
symbol we use the convention $\epsilon^{0123}=1$.\\ 
We will use the relativistic notation with repeated indices assumed to be 
summed over, however contractions of indices will be sometimes denoted with 
dots, e.g. $ u \cdot T \cdot u \equiv u_\mu T^{\mu\nu} u_\nu$. Operators in 
Hilbert space will be denoted by a large upper hat, e.g. $\wT$ while unit 
vectors with a small upper hat, e.g. $\hat v$. The stress-energy tensor is
assumed to be symmetric with an associated vanishing spin tensor.

\section{Equilibrium in relativistic quantum statistical mechanics}
\label{statmech}

A general covariant form of the density operator in relativistic quantum statistical
mechanics extending the eq.~(\ref{homo}) was first proposed, to our knowledge, in 
refs.~\cite{zubarev,weert}:
\be\label{gener1}
  \wrho = \frac{1}{Z} \exp \left[ -\int_{\Sigma} \di \Sigma_\mu \; 
  \left( \wT^{\mu\nu} \beta_\nu - \zeta \wj^\mu \right) \right]
\ee
where $\Sigma$ is a spacelike 3D hypersurface. This form can be obtained maximizing 
the total entropy with the constraints of given energy-momentum and charge densities 
at some specific "time" of the hypersurface $\Sigma$, see the detailed discussions 
in ref.~\cite{weert} and more recently in refs.~\cite{becalocal,japan}. The density
operator (\ref{gener1}) is therefore especially suitable to describe {\em local} 
thermodynamic equilibrium --- that is a situation where the thermodynamic parameters
temperature, velocity field and chemical potential are a function of space and time
--- in a quantum relativistic framework. The operator (\ref{gener1}) will not maintain
its form under the unitary time evolution and cannot thus represent {\em the} actual
quantum state in the Heisenberg representation. However, it is time independent or, 
equivalently, independent of the integration hypersurface $\Sigma$ if the divergence 
of the integrand vanishes and in this case the (\ref{gener1}) is the density operator
of a thermodynamic equilibrium state. For conserved stress-energy tensor and current
this condition leads to the request \cite{becacov} that $\zeta$ is a constant and 
$\beta$ a Killing vector field fulfilling eq.~(\ref{kill}) (with partial derivatives
replaced by covariant derivatives if necessary).

The density operator (\ref{gener1}) is also well suited to describe thermodynamic 
equilibrium in a general curved spacetime possessing a timelike Killing vector field. 
In Minkowski spacetime, which we will be dealing with in this work, the general 
solution of the eq.~(\ref{kill}) is:
\be\label{killsol}
   \beta^\nu = b^\nu + \varpi^{\nu\mu} x_\mu
\ee
where $b$ is a constant four-vector and $\varpi$ a constant antisymmetric tensor, 
which, because of (\ref{killsol}) can be written as an exterior derivative of the
$\beta$ field:
\be\label{thvort}
  \varpi_{\nu\mu} = -\frac{1}{2} (\partial_\nu \beta_\mu - \partial_\mu \beta_\nu)
\ee
Hence, the general equilibrium form in flat spacetime of the density operator (\ref{gener1})
reads:
\be\label{gener2}
  \wrho = \frac{1}{Z} \exp \left[ - b_\mu {\wP}^\mu  
  + \frac{1}{2} \varpi_{\mu\nu} \wJ^{\mu\nu} + \zeta \wQ \right]
\ee
where the $\wJ$'s are the generators of the Lorentz transformations:
$$
 \wJ^{\mu\nu} = \int_{\Sigma} \di \Sigma_\lambda \; \left( 
 x^\mu \wT^{\lambda\nu} - x^\nu \wT^{\lambda\mu} \right) 
$$
Therefore, besides the chemical potentials, the most general equilibrium density 
operator in Minkowski spacetime can be written as a linear combinations of the 10 
generators of the Poincar\'e group with 10 constant coefficients. The most widely 
known case is the one with $\beta = b$ and $\varpi=0$, that is eq.~(\ref{homo}),
what we define as {\em homogeneous thermodynamic equilibrium}. The rotating global 
equilibrium in eq.~(\ref{rotating}) can be obtained as a special case of eq.~(\ref{gener2}) 
setting:
$$
 b_\mu = (1/T_0,0,0,0) \qquad \qquad \varpi_{\mu\nu} = (\omega/T_0) (g_{1\mu} g_{2\nu} 
- g_{1\nu} g_{2\mu})
$$
where $\omega$ has the meaning of a costant angular velocity \cite{landau}. Similarly,
the form (\ref{accelerating}) can be obtained by setting:
$$
 b_\mu = (1/T_0,0,0,0) \qquad \qquad \varpi_{\mu\nu} = (a/T_0) (g_{0\mu} g_{3\nu}
 - g_{3\mu} g_{0\nu})
$$
In the latter case, the contravariant components of $\beta$ read: 
\be\label{rindler}
\beta^\mu= \frac{1}{T_0}\left(1 + a z, 0, 0,a t \right)
\ee
thus the unit vector $\hat \beta$ is the velocity field of a fluid with constant 
comoving acceleration along the field lines (for the field line going through $z=0$,
the comoving acceleration is $a$).

\section{Mean values of local operators}
\label{locality}

Suppose we want to calculate the mean value of a local operator $\widehat O(x)$ 
(in the Heisenberg picture) with the density operator (\ref{gener1}):
\be\label{lexpv}
 O(x) \equiv \langle \widehat O(x) \rangle = \tr (\wrho \, \widehat O(x))_{\rm ren} 
 =  \frac{1}{Z} \tr \left( \exp \left[ -\int_{\Sigma} \di \Sigma_\mu \; 
  \left( \wT^{\mu\nu} \beta_\nu - \zeta \wj^\mu \right) \right] \widehat O(x) 
  \right)_{\rm ren} \
\ee
If $\beta$ is a general field, there is no compelling reason why, at a given point 
$x$, the mean value $O(x)$ should be simply equal to the same value at the homogeneous 
global thermodynamic equilibrium with an uniform four-temperature $\beta$ equal 
to its value in the point $x$, that is $\beta(x)$. For instance, the stress-energy 
tensor in the point $x$ does not need to be of the ideal form (\ref{tideal}) 
with $u=\hat\beta(x)$ and $\rho = \rho(\beta^2,\zeta)$ $p = p(\beta^2,\zeta)$ 
if $\beta$ is not constant. In fact, its tensor structure in (\ref{tideal}) is 
determined by the symmetries of the density operator (\ref{homo}), which is obtained 
from (\ref{gener1}) provided that $\beta$ is constant.  

Nevertheless, one can imagine that if $\beta$ and $\zeta$ are sufficiently slowly 
varying in space and time, $O(x)$ will be mostly determined by the values of the fields 
$\beta$ and $\zeta$ around the point $x$ \cite{becalocal}. More specifically, the 
distance over which the {\em thermodynamic} fields like $\beta$ vary should be 
much larger than the typical thermal correlation length, which is governed by the 
microscopic parameters of the theory and the temperature itself. This can be shown 
by recasting the fields in the integrand of the eq.~(\ref{lexpv}) as $\beta = \beta(x) 
+ \delta \beta$ and $\zeta = \zeta(x) + \delta \zeta$, so as to obtain:
\begin{eqnarray*}
  O(x) &=& \frac{1}{Z} \tr \left( \exp \left[ - \beta_\nu(x) \int_{\Sigma} \di \Sigma_\mu \; 
  \wT^{\mu\nu} + \zeta(x) \int_{\Sigma} \di \Sigma_\mu \; \wj^{\mu} -
  \int_{\Sigma} \di \Sigma_\mu \; \left( \wT^{\mu\nu} \delta\beta_\nu - \delta \zeta 
  \wj^\mu \right) \right] \widehat O(x) \right)_{\rm ren} \nonumber \\
  &=& \frac{1}{Z} \tr \left( \exp \left[ - \beta_\nu (x) \wP^\nu + \zeta(x) \wQ 
  - \int_{\Sigma} \di \Sigma_\mu \; \left( \wT^{\mu\nu} \delta\beta_\nu - \delta \zeta 
  \wj^\mu \right) \right] \widehat O(x) \right)_{\rm ren}
\end{eqnarray*}
Hence, applying the linear response theory to the exponent above:
\bea\label{lrte}
  O(x) \simeq && \langle \widehat O (x) \rangle_{\beta(x)} - \int_0^1 \di z \; 
  \int_{\Sigma} \di \Sigma_\mu(y) \;  \left( \langle \widehat O (x) \wT^{\mu\nu}
  (y+iz\beta(x)) \rangle_{\beta(x)} - \langle \widehat O (x) \rangle_{\beta(x)} 
  \langle \wT^{\mu\nu}(y+iz\beta(x)) \rangle_{\beta(x)} \right) \delta\beta_\nu \nonumber \\  
  && + \int_0^1 \di z \; 
  \int_{\Sigma} \di \Sigma_\mu(y) \;  \left( \langle \widehat O (x) \wj^{\mu}
  (y+iz\beta(x)) \rangle_{\beta(x)} - \langle \widehat O (x) \rangle_{\beta(x)}
  \langle \wj^{\mu}(y+iz\beta(x)) \rangle_{\beta(x)} \right) \delta\zeta  
\eea
where $x$ and $y$ both lie on the hypersurface $\Sigma$. The symbol $\langle \; 
\rangle_{\beta}$ stands for the (renormalized) mean value calculated with the homogeneous 
equilibrium density operator in eq.~(\ref{homo}). Particularly, the $\langle \; \rangle_{\beta(x)}$
stands for the mean value calculated with a fixed four-temperature (and $\zeta$) equal to 
the value of the $\beta$ (and $\zeta$) fields in the point $x$. The formula (\ref{lrte})
just expresses the aforementioned concept, namely that the local equilibrium value 
of the operator $\widehat O(x)$ is determined by the local values of the thermodynamic 
fields with corrections depending on quantum-statistical correlations between operators 
in different points. These correlations -- hence the integrand function in eq.~(\ref{lrte}) 
-- are significant over microscopic distances $l$ dictated by the mass, temperature and 
coupling constants of the theory, which are supposedly much smaller than the macroscopic 
distance $L$ over which $\delta\beta$ and $\delta\zeta$ appreciably vary. This condition 
is usually referred to as hydrodynamical regime and in this regime terms beyond the linear 
in $\delta\beta$ and $\delta\zeta$ in eq.~(\ref{lrte}) contribute less and less as they 
are expected to be suppressed with higher powers of $l/L$.

Under these circumstances, it is possible to expand the thermodynamic fields in the 
eq.~(\ref{lexpv}) into a Taylor series about the point $x$. For a general $\beta$ 
field, this method makes it possible to find an approximate expression of the local 
thermodynamic equilibrium operator (\ref{gener1}) as a function of $\beta(x)$ and 
its derivatives \cite{becalocal}. For the special case of global equilibrium, with 
constant $\zeta$ and $\beta$ a Killing vector field (\ref{killsol}), one can recast 
the operator (\ref{gener2}) so as to have in the exponent the value of the four-temperature
in the point $x$:
\bea\label{geqmean}
 O(x) &=&  \frac{1}{Z} \tr \left( \exp \left[ - b_\mu {\wP}^\mu 
 + \frac{1}{2} \varpi_{\mu\nu} \wJ^{\mu\nu} + \zeta \wQ \right] \widehat 
 O(x) \right)_{\rm ren} \!\!\!\!\! = 
 \frac{1}{Z} \tr \left( \exp \left[ - (b_\mu + \varpi_{\mu\nu} x^\nu){\wP}^\mu 
 + \frac{1}{2} \varpi_{\mu\nu} \wJ^{\mu\nu}_x +\zeta \wQ \right] 
 \widehat O(x) \right)_{\rm ren} \nonumber \\
 &=& \frac{1}{Z} \tr \left( \exp \left[ - \beta_\mu(x) {\wP}^\mu 
 + \frac{1}{2} \varpi_{\mu\nu} \wJ^{\mu\nu}_x +\zeta \wQ \right] 
 \widehat O(x) \right)_{\rm ren}
\eea
where we have used the angular momentum operators around the point $x$:
\be\label{angmomtrasl}
 \wJ^{\mu\nu}_x = \wJ^{\mu\nu} - x^\mu \wP^\nu + x^\nu \wP^\mu = 
 \widehat{\sf T}(x) \wJ^{\mu\nu} \widehat{\sf T}(x)^{-1}
\ee
$\widehat{\sf T}(x) = \exp[\ii x \cdot \wP]$ being the translation operator.

The calculation of mean values (\ref{geqmean}) is the main purpose of this paper, 
and, specifically, when $\widehat O = \wT^{\mu\nu}$.
We will consider the term in $\varpi$ as small compared with the terms involving 
$\beta$ and $\zeta$ and expand accordingly. Thus, the leading term in the above 
equation will be simply the homogeneous equilibrium one with four-temperature equal 
to its value in the $x$ point, that is the expression (\ref{tideal}) with 
$u=\hat\beta(x)$. We will see in the sect.~\ref{expansion} that the lowest order 
corrections to the ideal form are of the second order in $\varpi$ and that they 
are of either quantum or quantum-relativistic nature as they vanish for $\hbar \to 0$ 
or $\hbar/c \to 0$.

Note that $\beta(x)$ is required to be a future-oriented timelike vector in order 
to get a finite value for most observables at the lowest order of the $\beta$ 
expansion. This condition cannot be fulfilled everywhere for the expression (\ref{killsol}) 
if $\varpi \ne 0$. For instance, for the rotating global equilibrium (\ref{rotating}), 
it is easy to check that:
$$
 \beta = \frac{1}{T_0} (1, \omega \hat{\bf k} \times {\bf x}) 
$$
which becomes spacelike when $ \| \omega \hat{\bf k} \times {\bf x} \| > 1$, that
is when the velocity exceeds the speed of light. Similarly, for the operator 
(\ref{accelerating}), the $\beta$ field (\ref{rindler}) is future-oriented timelike 
only in the Rindler wedge defined by the light cone of the point $(0,0,0,-a/T)$. 
Therefore, the validity of our calculations will be limited to the physical regions 
where the $\beta$ field is timelike and with positive time component even though the 
operator (\ref{gener2}) written with the constants $b$ and $\varpi$ does not make 
this limitation apparent.

\section{Perturbative expansion for the stress-energy tensor}
\label{expansion}

The goal of this section is to provide an expansion in $\varpi$ for the mean value of
the stress-energy tensor in the general form of thermodynamic equilibrium:
\be\label{setge}
 T^{\mu\nu}(x) = \frac{1}{Z} \tr \left( \exp \left[ - \beta_\mu(x) {\wP}^\mu 
 + \frac{1}{2} \varpi_{\mu\nu} \wJ^{\mu\nu}_x + \zeta \wQ \right] 
 \wT^{\mu\nu}(x) \right)_{\rm ren}
\ee
Indeed, $\varpi$ is an adimensional tensor in natural units and it the has, in general,
very small components. To understand its physical meaning, it is very useful to decompose
it into two spacelike vector fields, each having three independent components, projecting 
along a timelike vector. A physically interesting choice is $u = \hat\beta = \beta/\sqrt{\beta^2}$, 
in the regions where $\beta$ given by eq.~(\ref{killsol}) is timelike. We can then 
decompose $\varpi$ as follows:
\be\label{decomp1} 
  \varpi^{\mu\nu} = \epsilon^{\mu\nu\rho\sigma} w_\rho u_\sigma
  + \alpha^\mu u^\nu - \alpha^\nu u^\mu  
\ee
where, by definition:
\be\label{defin1}
  \alpha^\mu(x) = \varpi^{\mu\nu} u_\nu  \qquad \qquad w^\mu(x) = -\frac{1}{2}
  \epsilon^{\mu\nu\rho\sigma} \varpi_{\nu\rho} u_\sigma
\ee  
Note that $\alpha$ and $w$, unlike $\varpi$, are not constant and they are both 
orthogonal to $u$, hence spacelike. The physical meaning of $\alpha$ and $w$ vectors 
can be shown starting from the eq.~(\ref{kill}). Because of (\ref{killsol}) and 
(\ref{kill}) at equilibrium one has:
$$
\varpi_{\mu\nu} = \partial_\nu \beta_\mu
$$ 
whence:
$$
  \alpha^\mu = \varpi^{\mu\nu} u_\nu = u_\nu \partial^\nu \beta^\mu 
  = u^\mu u_\nu \partial^\nu \sqrt{\beta^2} + \sqrt{\beta^2} 
  u_\nu \partial^\nu u^\mu 
$$
We can now take the scalar product with $u^\mu$ and conclude that:
$$
  u_\nu \partial^\nu \sqrt{\beta^2} \equiv D\sqrt{\beta^2} = 0 
$$
which tells us that, as expected, at the thermodynamic equilibrium the comoving
temperature along the flow lines does not change and $\partial_\mu \beta^2 = 
\nabla_\mu \beta^2$, where:
$$
  \nabla_\mu \equiv \partial_\mu - u_\mu D
$$
Thereby, the $\alpha$ vector simply becomes:
\be\label{accel}
  \alpha^\mu = \sqrt{\beta^2} u_\nu \partial^\nu u^\mu = \sqrt{\beta^2} Du^\mu 
  = \frac{1}{T} a^\mu
\ee
that is the acceleration field divided by the proper temperature. Note also that,
being $\partial_\mu \beta_\nu + \partial_\nu \beta_\mu = 0$, one has:
\be\label{accel2}
  0 = u^\nu (\partial_\nu \beta_\mu + \partial_\mu \beta_\nu) = 
  \alpha_\mu + \frac{1}{2 \sqrt{\beta^2}} \partial_\mu \beta^2 = \frac{1}{T} 
  a_\mu - \frac{1}{T^2} \nabla_\mu T
\ee
Likewise, it can be shown that $w$ corresponds to an angular velocity divided by a 
temperature, for, by using (\ref{thvort})
\be\label{angvel}
  w^\mu = -\frac{1}{2} \epsilon^{\mu\nu\rho\sigma} \varpi_{\nu\rho} u_\sigma
  = \frac{1}{2} \epsilon^{\mu\nu\rho\sigma} (\partial_\nu \beta_\rho) u_\sigma
  = \frac{1}{2} \epsilon^{\mu\nu\rho\sigma} \sqrt{\beta^2} u_\sigma \partial_\nu u_\rho
  = \frac{1}{2T} \epsilon^{\mu\nu\rho\sigma} u_\sigma \nabla_\nu u_\rho
  = \frac{1}{T} \omega^\mu
\ee
being $\omega^\mu = \frac{1}{2} \epsilon^{\mu\nu\rho\sigma} u_\sigma \nabla_\nu u_\rho$,
as it is known in literature, the local vorticity vector. Restoring the physical 
constants, one then has the adimensional four-vectors:
\be\label{acceler} 
  \alpha_\mu = \frac{\hbar a_\mu}{c K T}  \qquad \qquad w_\mu = \frac{\hbar 
  \omega_\mu}{K T} 
\ee
These numbers are, for the vast majority of physical systems, much less than 1 and
a perturbative expansion in $\varpi$ of the eq.~(\ref{setge}) is then feasible. 
They can give rise to relevant corrections if the implied additional terms to the
ideal stress-energy tensor are some sizeable fraction thereof or when these terms 
are comparable to the viscous tensor. According to the eq.~(\ref{acceler}), this happens 
at for very large accelerations or very low temperatures. 
Hence, let us define:
\be\label{wrdef}
 \wR(\varpi) \equiv \exp \left[ -\beta_\mu(x) {\wP}^\mu + \frac{1}{2} 
 \varpi_{\mu\nu} \wJ^{\mu\nu}_x + \zeta \wQ \right] = 
  \exp \left[ -\beta_\mu(x) {\wP}^\mu + \frac{1}{2} \varpi_{\mu\nu} 
  \wJ^{\mu\nu}_x \right] \exp[\zeta \wQ] 
\ee
where, in the last equality, advantage has been taken of the supposed commutation 
of the charge operator $\wQ$ with both the $\wP$'s and $\wJ$'s. At the second order in 
$\varpi$ one can write:
\be\label{expand}
 \wR(\varpi) = \wR^{(0)} + \varpi_{\mu\nu} \wR^{(1)\mu\nu}
 + \varpi_{\mu\nu}\varpi_{\rho\sigma} \wR^{(2)\mu\nu\rho\sigma} +
 o(\varpi^2) 
\ee
and, by using the Poincar\'e group commutation relations, it can be shown that (see 
Appendix A):
\bea\label{expand2}
 \wR^{(0)}  &=& \e^{-\beta \cdot \wP + \zeta \wQ} \nonumber \\
 \wR^{(1)\mu\nu}  &=& \frac{1}{4} \{ \e^{-\beta \cdot \wP + \zeta \wQ} , 
 \wJ^{\mu\nu} \} \, , \nonumber \\
 \wR^{(2)\mu\nu\rho\sigma} &=& \frac{1}{16} \{ \e^{-\beta \cdot \wP 
 + \zeta \wQ},\wJ^{\mu\nu} 
 \wJ^{\rho\sigma} \} + \frac{1}{8} \e^{-\beta \cdot \wP + \zeta \wQ} \beta^\mu \beta^\rho 
 \wP^\nu \wP^\sigma  -\frac{1}{12} \e^{-\beta \cdot \wP + \zeta \wQ} \beta^\mu g^{\nu\rho} 
 \wP^\sigma \, .
\eea
where the curly bracket expression $\{\; , \; \}$ stands for the anticommutator.

By using the eqs.~(\ref{expand}) and (\ref{expand2}), the mean value (\ref{setge}) 
can be expressed as an expansion in $\varpi$ with coefficients which are calculated 
at the homogeneous thermodynamic equilibrium:
\bea\label{texpa}
 T^{\mu\nu}(x)= && \frac{\tr(\wR(\varpi)\wT^{\mu\nu}(x))}{\tr(\wR(\varpi))} =
 \langle \wT^{\mu\nu}(x) \rangle_{\beta(x)} + \frac{1}{2} \varpi_{\rho\sigma} 
 {\rm Re} \langle \wJ^{\rho\sigma}_x ; \wT^{\mu\nu}(x) \rangle_{\beta(x)} \nonumber \\
 && + \varpi_{\rho\sigma} \varpi_{\lambda\tau} 
 \left[ \frac{1}{8} {\rm Re} \langle \wJ^{\rho\sigma}_x \wJ^{\lambda\tau}_x ; 
 \wT^{\mu\nu}(x) \rangle_{\beta(x)} + \frac{1}{8} \beta^\rho(x) \beta^\lambda(x)
 \langle \wP^{\sigma} \wP^{\tau} ; \wT^{\mu\nu}(x) \rangle_{\beta(x)} - 
 \frac{1}{12} \beta^\rho(x) g^{\lambda\sigma} \langle \wP^{\tau} ; \wT^{\mu\nu}(x) 
 \rangle_{\beta(x)} \right. \nonumber \\
 && \left. - \frac{1}{4} {\rm Re} \langle \wJ^{\rho\sigma}_x ; \wT^{\mu\nu}(x) 
 \rangle_{\beta(x)} \langle \wJ^{\lambda\tau}_x \rangle_{\beta(x)} \right] + o(\varpi^2)
\eea
where we have used the relations for two hermitian operators $\wA,\wB$:
$$
 {\rm Re} \langle \wA \wB \rangle = \frac{1}{2} \langle \{\wA,\wB\} \rangle
 \qquad \qquad
 i {\rm Im} \langle \wA \wB \rangle = \frac{1}{2} \langle [\wA,\wB] \rangle
$$
and the notation has been introduced:
$$
 \langle \widehat A; \widehat B \rangle = \langle \widehat A \widehat B 
 \rangle - \langle \widehat A \rangle \langle \widehat B \rangle
$$
for the correlator between $\wA$ and $\wB$. The terms in eq.~(\ref{texpa}) containing 
$\wP$ can be readily calculated taking the derivative of $\langle \wT \rangle_\beta$ 
with respect to $\beta$. Indeed:
\bea\label{tderivat}
 \beta^\rho g^{\lambda\sigma} \frac{\partial}{\partial \beta_\tau} \langle \wT^{\mu\nu}
 \rangle_\beta &=& -\beta^\rho g^{\lambda\sigma} \langle \wP^\tau ; \wT^{\mu\nu} \rangle_\beta
 \nonumber \\
  \beta^\rho \beta^\lambda \frac{\partial^2}{\partial\beta_\sigma \,\partial\beta_\tau} 
  \langle \wT^{\mu\nu} \rangle_\beta &=& \beta^\rho \beta^\lambda \left(
  \langle \wP^\sigma \wP^\tau ; \wT^{\mu\nu} \rangle_{\beta} -
  \langle \wP^\sigma ; \wT^{\mu\nu} \rangle_{\beta} \langle \wP^\tau \rangle_{\beta}
 - \langle \wP^\tau ; \wT^{\mu\nu} \rangle_{\beta} \langle \wP^\sigma \rangle_{\beta} \right)
\eea
Note that the two rightmost terms in the last equation vanish once multiplied by 
$\varpi_{\rho\sigma}\varpi_{\lambda\tau}$ for, being $\langle \wP \rangle_\beta
\propto \beta$, they contain the symmetric combination $\beta^\lambda \beta^\tau$ or 
$\beta^\rho \beta^\sigma$. 

All mean values in eq.~(\ref{texpa}) involving angular momentum operators can be 
rewritten in a form which makes it apparent that their dependence on $x$ is only
through the value of the four-temperature, by taking advantage of the translational 
invariance of the density operator. For instance:
$$
 \langle \wJ^{\rho\sigma}_x \wJ^{\lambda\tau}_x ; \wT^{\mu\nu}(x) \rangle_{\beta(x)}
 = \langle \widehat{\sf T}^{-1}(x) \wJ^{\rho\sigma}_x \wJ^{\lambda\tau}_x \widehat{\sf T}(x); 
 \widehat{\sf T}^{-1}(x) \wT^{\mu\nu}(x) \widehat{\sf T}(x) \rangle_{\beta(x)} = 
 \langle \wJ^{\rho\sigma} \wJ^{\lambda\tau} ; \wT^{\mu\nu}(0) \rangle_{\beta(x)} 
$$
and similarly for the others, where eq.~(\ref{angmomtrasl}) has been used. Then, 
it is convenient to decompose the tensor $\wJ$ into two spacelike vector operators 
the same fashion as for $\varpi$ in eq.~(\ref{decomp1}):
\be\label{decomp2}
 \wJ^{\mu\nu} = u^\mu \wK^{\nu} - \wK^{\mu} u^\nu + 
 \epsilon^{\mu\nu\rho\sigma} \wJ_{\rho} u_\sigma
\ee 
being
$$
\wK^\mu = u_\rho \wJ^{\rho\mu} \qquad \wJ^\mu = - \frac{1}{2} \epsilon^{\mu\rho\sigma\tau} 
\wJ_{\rho\sigma} u_\tau
$$
The operators $\wJ^\mu$ and $\wK^\mu$ are simply the generators of the rotation 
and boosts with respect to the reference frame with time direction $u$. Using the
invariance by rotation (in the hyperplane orthogonal to $u$), parity and time 
reversal, which are assumed to hold for our hamiltonian, one readily obtains that 
(see Appendix B):
$$
 {\rm Re} \langle \wJ^{\rho\sigma} ; \wT^{\mu\nu}(0) \rangle_{\beta(x)} = 0 \qquad
 \langle \wJ^{\rho\sigma} \rangle_{\beta(x)} = 0 
$$
Therefore, plugging the decomposition (\ref{decomp2}) into the eq.~(\ref{texpa}), 
and using the relations (\ref{defin1}) and (\ref{tderivat}) and after the removal
of the vanishing terms, the eq.~(\ref{texpa}) can be written as:
\bea\label{texpa2}
 T^{\mu\nu}(x) &=&  
 \langle \wT^{\mu\nu}(x) \rangle_{\beta(x)} + \frac{1}{2}  \alpha_\rho \alpha_\sigma
 {\rm Re} \langle \wK^{\rho} \wK^{\sigma} ; \wT^{\mu\nu}(0) \rangle_{\beta(x)}
 + \frac{1}{2} w_\rho w_\sigma {\rm Re} \langle \wJ^{\rho} \wJ^{\sigma} ; 
 \wT^{\mu\nu}(0) \rangle_{\beta(x)} \nonumber \\
 &+&  \frac{1}{2} \alpha_\rho w_\sigma {\rm Re} \langle \{ \wJ^{\rho}, \wK^{\sigma} \} 
 ; \wT^{\mu\nu}(0) \rangle_{\beta(x)} + \frac{1}{8} \beta^\rho \beta^\lambda
 \frac{\partial^2}{\partial\beta_\sigma \,\partial\beta_\tau} \langle \wT^{\mu\nu}(x) 
 \rangle_{\beta(x)} + \frac{1}{12} \beta^\rho g^{\lambda\sigma} 
 \frac{\partial}{\partial \beta_\tau} \langle \wT^{\mu\nu}(x) \rangle_{\beta(x)}
 + o(\varpi^2)
\eea
The derivative terms are easy to work out by using (\ref{tideal}); they will give 
rise to expressions involving the thermodynamic functions pressure, energy density
and their derivatives, that is specific heats. On the other hand, the correlators 
in eq.~(\ref{texpa2}) cannot be expressed in terms of known thermodynamic functions. 
In fact, they can be written as linear combinations of new thermodynamic coefficients 
which can be expressed in turn as correlators of specific components of the stress-energy 
tensor and angular momentum or boost operators $\wJ$ and $\wK$, that is
\bea\label{corrcoeff}
 && \frac{1}{2} {\rm Re} \langle \{ \wK^{\rho}, \wK^{\sigma}\} ; \wT^{\mu\nu}(0) \rangle_{\beta(x)}
  = - u^\mu u^\nu \Delta^{\rho\sigma} k_t(T,\zeta) +  \Delta^{\mu\nu} \Delta^{\rho\sigma} k_\theta(T,\zeta) 
  + (\Delta^{\mu\sigma} \Delta^{\rho\nu} + \Delta^{\nu\sigma} \Delta^{\rho\mu} 
  - \frac{2}{3}\Delta^{\mu\nu}\Delta^{\rho\sigma}) k_s(T,\zeta)   \nonumber \\
 && \frac{1}{2} {\rm Re} \langle \{ \wJ^{\rho}, \wJ^{\sigma} \} ; \wT^{\mu\nu}(0) \rangle_{\beta(x)}
  = - u^\mu u^\nu \Delta^{\rho\sigma} j_t(T,\zeta) +  \Delta^{\mu\nu} \Delta^{\rho\sigma} j_\theta(T,\zeta) 
  + (\Delta^{\mu\sigma} \Delta^{\rho\nu} + \Delta^{\nu\sigma} \Delta^{\rho\mu} 
  - \frac{2}{3}\Delta^{\mu\nu}\Delta^{\rho\sigma}) j_s(T,\zeta) \nonumber \\
 && {\rm Re} \langle \{ \wK^{\rho}, \wJ^{\sigma} \} ; \wT^{\mu\nu}(0) \rangle_{\beta(x)}
  = (u_\mu u_\kappa \epsilon^{\kappa\rho\sigma\nu} + u^\nu u_\kappa \epsilon^{\kappa\rho\sigma\mu})l_v(T,\zeta)  
\eea
where
$$
 \Delta^{\mu\nu} \equiv g^{\mu\nu} - u^\mu u^\nu
$$
is the projector onto the hyperplane orthogonal to $u$ and
\begin{align}\label{correlat}
  k_t(T,\zeta)      &= {\rm Re}\langle \wK^{3} \wK^{3}; \wT^{00}(0) \rangle_T 
  \qquad &
  k_\theta(T,\zeta) &= \frac{1}{3} {\rm Re} \sum_{i=1}^3 \langle \wK^{3} \wK^{3}; 
   \wT^{ii}(0) \rangle_T \qquad &
  k_s(T,\zeta)      &= {\rm Re} \langle \wK^{1} \wK^{2}; \wT^{12}(0) \rangle_T \, \nonumber \\ 
  j_t(T,\zeta)      &= {\rm Re} \langle \wJ^{3} \wJ^{3}; \wT^{00}(0) \rangle_T 
  \qquad &
  j_\theta(T,\zeta) &= \frac{1}{3} {\rm Re} \sum_{i=1}^3 \langle \wJ^{3} \wJ^{3}; \wT^{ii}(0) \rangle_T 
  \qquad &
  j_s(T,\zeta)      &= {\rm Re}\langle \wJ^{1} \wJ^{2}; \wT^{12}(0) \rangle_T \, \nonumber \\
  l_v(T,\zeta)      &= {\rm Re}\langle \{\wK^1,\wJ^2 \}; \wT^{03}(0) \rangle_T 
\end{align}
In the eq.~(\ref{correlat}) the notation $\langle \; \rangle_T$ has been introduced meaning
that the expectation value is calculated in the rest frame where $\beta=(1/T,{\bf 0})$.
The derivation of eqs.~(\ref{corrcoeff}) and (\ref{correlat}) can be found in Appendix B.

Finally, after having worked out the derivatives of the stress-energy tensor and using
the eqs.~(\ref{corrcoeff}) and (\ref{tideal}) in the eq.~(\ref{texpa2}), one obtains:
\vspace*{0.5cm}
\be\label{texpa3}
 T^{\mu\nu}(x) = (\rho - \alpha^2 U_\alpha- w^2 U_w) 
 u^\mu u^\nu -  (p - \alpha^2 D_\alpha - w^2 D_w) \Delta^{\mu\nu} + A \alpha^\mu \alpha^\nu 
 + W w^\mu w^\nu + G (u^\mu \gamma^\nu + \gamma^\mu u^\nu) +o(\varpi^2)
\vspace*{0.5cm}
\ee
where $\rho,p$ are the usual homogeneous thermodynamic equilibrium functions energy
density and pressure, and the functions $U,D,A,W,G$ read:
\begin{align}\label{udaw}
 U_\alpha &= \frac{1}{24} T \frac{\partial \rho}{\partial T} 
 + \frac{1}{4}(\rho+p) + \frac{1}{2} k_t \qquad\qquad & U_w &= \frac{1}{2}j_t  \qquad\qquad 
 & D_\alpha &= \frac{1}{24}(\rho+p) +\frac{1}{2}k_\theta - \frac{1}{3}k_s \nonumber \\
 D_w &= \frac{1}{2}j_\theta - \frac{1}{3}j_s \qquad\qquad &
 A &= \frac{1}{4}(\rho+p)+k_s  \qquad\qquad & W &= j_s \nonumber \\
 G &= \frac{1}{2} l_v - \frac{1}{12}(\rho+p)
\end{align}
The vector $\gamma^\mu$ in eq.~(\ref{texpa3}) is defined as:
\be\label{gamma}
 \gamma^\mu = (\alpha\cdot\varpi)_\lambda \Delta^{\lambda\mu} = 
 \epsilon^{\mu\nu\rho\sigma} w_\nu \alpha_\rho u_\sigma  
\ee
where the $\varpi$ decomposition (\ref{decomp1}) has been used,

As it can be seen from eq.~(\ref{texpa3}), the stress-energy tensor has corrections
to its ideal form which depend on quadratic combinations of the two vector fields,
$\alpha$ and $w$ arising from the decomposition of the exterior derivative of the
temperature four-vector $\beta$. At thermodynamic equilibrium, according to the 
previous discussion and the eqs.~(\ref{acceler}), they are proportional to the 
acceleration $a^\mu$ and angular velocity (or vorticity) $\omega^\mu$, so that the 
eq.~(\ref{texpa3}) can be rewritten in the most suggestive fashion by restoring the 
natural constants as: 
\bea\label{texpa4}
  T^{\mu\nu}(x) = && \left[ \rho + \left(\frac{\hbar|a|}{c KT}\right)^2 
 U_\alpha + \left(\frac{\hbar |\omega|}{KT}\right)^2  U_w \right] u^\mu u^\nu -  
 \left[ p +  \left(\frac{\hbar |a|}{c KT}\right)^2  D_\alpha + \left(\frac{\hbar |\omega|}{KT}\right)^2 
 D_w \right] \Delta^{\mu\nu} \nonumber \\
 && + A \left(\frac{\hbar |a|}{c KT}\right)^2 \hat a^\mu \hat a^\nu + 
  W \left(\frac{\hbar |\omega|}{KT}\right)^2 \hat\omega^\mu \hat\omega^\nu +  
  G \frac{\hbar^2 |\omega| |a|}{c(KT)^2} (u^\mu \hat\gamma^\nu + \hat\gamma^\mu u^\nu)
  + o(\varpi^2)
\eea
where $|a| = \sqrt{-a_\mu a^\mu}$ and $|\omega| = \sqrt{-\omega_\mu \omega^\mu}$
and $\hat a$, $\hat \omega$ are the corresponding unit vectors. 
In the expression (\ref{texpa4}) the adimensional scales $\hbar a/cKT$ and $\hbar\omega/KT$
involving acceleration and vorticity have been separated from the thermodynamic 
functions $U,A,W,D,G$ having the same dimension as $\rho$ and/or $p$ making it 
easier to appreciate the size of the correction to the ideal form.

\subsection{Relation with other second-order hydrodynamical coefficient calculations}

The appearance of extra terms in the stress-energy tensor at thermodynamic equilibrium 
with respect to its ideal form has, needless to say, several physical consequences. 
The presence of non-dissipative quadratic corrections in the vorticity and gradients 
of temperature (hence accelerations at equilibrium, according to eq.~(\ref{accel2})) 
was pointed out in ref.~\cite{roma1,roma2} and, being non dissipative in nature, defined 
as thermodynamic in ref.~\cite{moore}. The calculation of such coefficients has 
attracted much attention lately (see \cite{kodama} for a recent review), especially 
in conformal field theories \cite{arnold,starinets,brazil} with different techniques 
\cite{molnar,philipsen} (see also ref.~\cite{jaiswal}). The coefficients that we have
denoted by $D_\alpha$, $D_w$, $A$ and $W$ are in the following relation with those 
known as $\xi_3,\xi_4,\lambda_3,\lambda_4$ in literature:
\begin{align}\label{corresp}
 \frac{A}{T^2}   &= 9 \lambda_4  \qquad \qquad  & \frac{W}{T^2} &= \lambda_3  \nonumber \\
 \frac{D_w}{T^2} &= \left( \frac{\lambda_3}{3} - 2 \xi_3 \right) \qquad \qquad &
  \frac{D_\alpha}{T^2} &= \left( 3\lambda_4 - 9 \xi_4 \right)  
\end{align}
Remarkably, the number of coefficients quoted in (\ref{texpa4}) is larger than envisaged
in ref.~\cite{roma1,roma2} and the reason is that we did not assume, as it is usually
done in the Landau frame, that the proper energy density $\rho$ has the same 
functional dependence on the temperature as at homogeneous thermodynamic equilibrium. 
This assumption proves to be incorrect, and the extra coefficients cannot be reabsorbed
by a redefinition of temperature, as it will be discussed and shown in sect.~\ref{discuss}.

Before tackling these issues, it is necessary to calculate the coefficients $U, D, A, W, G$
in some instance and we will do it for the simplest case of a real scalar free field. 
As it will be clear from the calculations shown in the next section, they all have 
a classical expression in the massive case, and, as a consequence, all the corrections 
in eq.~(\ref{texpa4}) to the ideal form turn out to be of quantum origin, as they vanish 
in the $\hbar \to 0$ limit.

\section{The free scalar field}
\label{freef}

The goal of this section is to calculate the coefficients in eq.~(\ref{texpa4}) for 
a free real scalar field. This implies $\zeta = 0$ in the density operator (\ref{homo}),
yet it is quite easy to extend the obtained results to the charged case with $\zeta\ne 0$
in the Boltzmann limit of distinguishable particles.

The theory is described by the Lagrangian density:
$$
\mathcal{L} = \frac{1}{2} \partial_\mu \wpsi \, \partial^\mu \wpsi - \frac{1}{2} m^2 \wpsi^2 \, .
$$
By adding the super-potential
$$
-2\xi\partial_\mu(\wpsi \, \partial^\mu\wpsi)
$$
a class of stress-energy tensors can be obtained as Noether currents associated to 
space-time translations. Although they are explicitely dependent on the parameter
$\xi$, they differ from each other by a divergence:
\be\label{scalart}                 
\wT_\xi^{\mu\nu} = \; \partial^\mu \wpsi \partial^\nu \wpsi - \frac{1}{2} g^{\mu\nu}  
 \left( \partial_\lambda \wpsi \, \partial^\lambda\wpsi - m^2\wpsi^2 \right) +
 2 \xi \; \partial_\lambda \left( g^{\mu\nu} \wpsi \, \partial^\lambda \wpsi -
  g^{\lambda\mu} \wpsi \, \partial^\nu \wpsi \right) 
\ee
thus they lead to the same generators of the Poincar\'e group. For $\xi=0$ the tensor 
is the so-called \emph{canonical stress-energy tensor}, while for $\xi=1/6$ 
the tensor is the so-called \emph{improved stress-energy tensor} \cite{callan}.  
In the translationally invariant homogeneous equilibrium (\ref{homo}) all mean values 
of local operators are independent of $x$, thus the divergence in the above expression 
vanishes, hence $\rho$ and $p$ do not depend on $\xi$. In fact, as we will show, 
this is not true in the case of generalized equilibrium and the correlators in 
eq.~(\ref{correlat}) are explicitely dependent on $\xi$.

At the very beginning, it should be pointed out that in principle one should use normal
ordering in the calculation of the mean values of $\wT$ in a free field theory to 
subtract zero point infinity. However, this is not needed in the calculation of a 
correlator such as $\langle \wJ^{\mu\nu} \wJ^{\rho\sigma} ; \wT^{\alpha\beta}(0) 
\rangle_T$ because $:\!\wT\!: = \wT - \bra{0} \wT \ket{0}$ ($\wT$ being a quadratic operator 
in the fields) so that the vacuum term cancels out in the subtraction $\langle \wJ^{\mu\nu} 
\wJ^{\rho\sigma} \wT^{\alpha\beta}(0) \rangle_T - \langle \wJ^{\mu\nu} \wJ^{\rho\sigma} 
\rangle_T \langle \wT^{\alpha\beta}(0) \rangle_T$.

The basic tool we need in order to carry out the calculation is the free field 
$n$-points Wightman thermal function:
$$
\mathcal{W}^{(n)}_T(x_1,x_2, \ldots, x_n) = \langle\wpsi(x_1)\,
\wpsi(x_2)\,\ldots\,\wpsi(x_n)\rangle_T
$$
which can be written in terms of 2-points thermal functions according to a version 
of the Wick theorem \cite{evans} suitable for thermal field theory. For an even $n$:
$$
\mathcal{W}^{(n)}_T(x_1,\dots ,x_n) = \sum_{j=2}^n \Big[ \mathcal{W}^{(2)}_T(x_1,x_j) 
\mathcal{W}^{(n-2)}_T(x_2,\dots ,x_{j-1},x_{j+1},\dots ,x_n)\,\Big] \, ,
$$
while, if $n$ is odd, $\mathcal{W}^{(n)}_T(x_1,\dots ,x_n) = 0$.
In the case of a free real scalar field, the 2-points Wightman thermal function 
reads:
$$
\mathcal{W}^{(2)}_T(x,y) = \frac{1}{(2\pi )^3} \int \mathrm{d}^4k \, e^{-ik(x-y)} 
\left[ \theta (k^0) + n_T(|k^0|) \right] \delta(k^2-m^2) \, .
$$
where
$$
n_T(\varepsilon ) = \frac{1}{\e^{\varepsilon/T} - 1}
$$
is the Bose--Einstein distribution. We then define
$$
\mathcal{T}_T^{\mu\nu |\rho\sigma |\alpha\beta}(x,y,z) = \; \langle 
\wT^{\mu\nu}(x) \, \wT^{\rho\sigma}(y) \, \wT^{\alpha\beta}(z)\rangle_T 
- \langle \wT^{\mu\nu}(x) \, \wT^{\rho\sigma}(y)\rangle_T \, 
\langle \wT^{\alpha\beta}(z)\rangle_T \, ,
$$
which can be calculated with the point-split procedure as
$$
\mathcal{T}_T^{\mu\nu |\rho\sigma |\alpha\beta}(x,y,z) 
= \Theta^{\mu\nu}_x \, \Theta^{\rho\sigma}_y \, 
\Theta^{\alpha\beta}_z \Big[\mathcal{W}^{(6)}_T(x_1,x_2,y_1,y_2,z_1,z_2) 
- \mathcal{W}^{(4)}_T(x_1,x_2,y_1,y_2)\, \mathcal{W}^{(2)}_T(z_1,z_2)\Big] \, ,
$$
where
$$
\Theta^{\mu\nu}_x = \big\{ (1-2\xi) \partial_{x_1}^\mu  
\partial_{x_2}^\nu -2\xi \partial_{x_2}^\mu  \partial_{x_2}^\nu 
+ \frac{1}{2} g^{\mu\nu} \big[ (4\xi - 1) \partial^{\vphantom{\mu}}_{x_1} 
\!\!\cdot \partial^{\vphantom{\mu}}_{x_2} + 4\xi\square^{\vphantom{\mu}}_{x_2} 
+ m^2 \big] \big\}_{x_1,x_2\rightarrow x} \, .
$$
The general expression of the correlators is then:
\bea\label{JJT}
\langle \wJ^{\mu\nu} \wJ^{\rho\sigma} ; \wT^{\alpha\beta}(0) \rangle_T = &&
\int \di^3 \x \, \di^3 \y \big[ x^\mu y^\rho \mathcal{T}_T^{0\nu |0\sigma |\alpha\beta}(x,y,0)
 - x^\nu y^\rho \mathcal{T}_T^{0\mu |0\sigma |\alpha\beta}(x,y,0) \nonumber \\
 && - x^\mu y^\sigma \mathcal{T}_T^{0\nu |0\rho |\alpha\beta}(x,y,0) + 
x^\nu y^\sigma \mathcal{T}_T^{0\mu |0\rho |\alpha\beta}(x,y,0)\big] \, 
\eea
with $x^0=y^0=0$ because the $\wJ$'s are time-independent.

Out of the 15 different diagrams stemming from the contractions of the 6-point Wightman 
thermal function, in $\mathcal{T}$ some are cancelled by the subtraction term, leaving 
only the 12 diagrams in which $\wT^{\alpha\beta}(z)$ is not a disconnected component. Since 
$\langle \wJ^{\mu\nu} \rangle_T=0$, in the eq.~(\ref{JJT}) the remaining 4 disconnected 
graphs in $\mathcal{T}$ do not contribute to the result. Therefore in (\ref{JJT}) we 
can replace $\mathcal{T}$ with its connected subset of 8 diagrams and we get:
\begin{figure}
$$
\mathcal{C}_T^{\mu\nu |\rho\sigma |\alpha\beta}(x,y,z) = \;
\raisebox{-.4\height}{\includegraphics[keepaspectratio=true,scale=1]{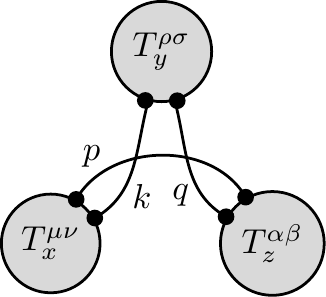}} \; + \;
\raisebox{-.4\height}{\includegraphics[keepaspectratio=true,scale=1]{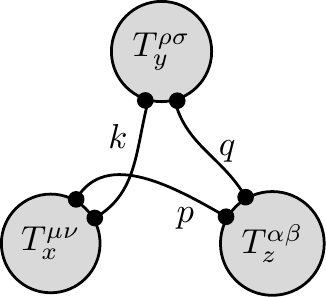}} \; + \;
\raisebox{-.4\height}{\includegraphics[keepaspectratio=true,scale=1]{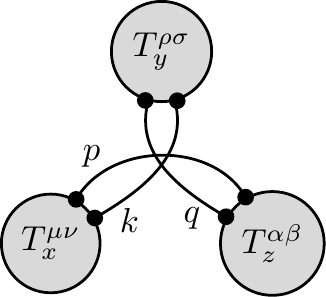}} \; + \; \ldots \, . 
$$
\end{figure}
\begin{multline*}
\mathcal{C}_T^{\mu\nu |\rho\sigma |\alpha\beta}(x,y,0) = \frac{1}{(2\pi )^9}  \int  
\mathrm{d}^4k\, \mathrm{d}^4p\, \mathrm{d}^4q\,\e^{-\ii (k+p)x}\,\e^{-\ii (q-k)y}\,
\mathcal{P}^{\mu\nu|\rho\sigma|\alpha\beta}(k,p,q) \, \delta (k^2-m^2)\,\delta (p^2-m^2)\,
\delta (q^2-m^2)
\\ \times \left[\theta (k^0)+n_T(|k^0|)\right] \left[\theta (p^0)+n_T(|p^0|)\right] 
\left[\theta (q^0)+n_T(|q^0|)\right] \, ,
\end{multline*}
with
\begin{equation*}
\begin{aligned}
\mathcal{P}^{\mu\nu|\rho\sigma|\alpha\beta}(k,p,q) = &\left\{-(1-2\xi)(k^\mu p^\nu + p^\mu k^\nu) 
+ 2\xi (k^\mu k^\nu + p^\mu p^\nu) - g^{\mu\nu} \left[(4\xi - 1) k \cdot p + 
2\xi(k^2+p^2)-m^2\right]\right\} \\
&\times \left\{(1-2\xi)(k^\rho q^\sigma + q^\rho k^\sigma) + 2\xi (k^\rho k^\sigma + 
q^\rho q^\sigma) - g^{\rho\sigma} \left[-(4\xi - 1) k \cdot q + 2\xi(k^2+q^2)-m^2\right]
\right\} \\
&\quad\times \left\{-(1-2\xi)(p^\alpha q^\beta + q^\alpha p^\beta) + 2\xi 
(p^\alpha p^\beta + q^\alpha q^\beta) - g^{\alpha\beta} \left[(4\xi - 1) p \cdot q + 
2\xi(p^2+q^2)-m^2\right]\right\} \, .
\end{aligned}
\end{equation*}

The thermodynamic correlators in eq.~(\ref{correlat}) can be found by selecting the 
suitable indices in eq.~(\ref{JJT}). For instance, for the $k_t$ correlator:
\begin{equation}\label{kt_starting_point}
 k_t(T) =  \int \di^3 \x \, \di^3 \y \; x^3 y^3 \, 
 \mathcal{C}_T^{00|00|00}(x,y,0)\Big|_{x^0=y^0=0} \, .
\end{equation}
Using
$$
\int \di^3 \x \, \di^3 \y \; x^i y^j
 \e^{\ii (\mathbf{k}+\mathbf{p})\cdot\mathbf{x}}\e^{i(\mathbf{q}-\mathbf{k})\cdot\mathbf{y}} 
 = -(2\pi)^6 \, \partial_{p_i}\delta(\mathbf{p}-\mathbf{k}) \, 
 \partial_{q_j}\delta(\mathbf{q}-\mathbf{k})
$$
and
$$
\delta (k^2-m^2) = \frac{1}{2\varepsilon_{\mathbf{k}}} \left[\delta (k^0+\varepsilon_{\mathbf{k}}) 
+ \delta (k^0-\varepsilon_{\mathbf{k}})\right] \, ,
$$
where $\varepsilon_{\mathbf{k}}=\sqrt{\mathbf{k}^2+m^2}$, one can then integrate in $\mathbf{x}$ 
and $\mathbf{y}$, thereafter in $k^0$, $p^0$, $q^0$ so as to obtain
$$
 k_t(T) = -\frac{1}{(2\pi)^3} \int \di^3 {\rm k} \, \di^3 \p \, 
 \di^3 {\rm q} \; \frac{1}{8\varepsilon_\mathbf{k}\varepsilon_\mathbf{p}\varepsilon_\mathbf{q}} 
 \left(\mathcal{S}_{+++}+\cdots+\mathcal{S}_{---}\right) \, \partial_{p_3}\delta(\mathbf{p}-\mathbf{k}) 
 \, \partial_{q_3}\delta(\mathbf{q}-\mathbf{k}) \, ,
$$
where the $\mathcal{S}$ terms correspond to the 8 possible combinations of positive 
and negative frequency of the $k$, $p$ and $q$ four-momenta. Thus, we have
\begin{eqnarray*}
\mathcal{S}_{+++} &=& \mathcal{P}_T^{00|00|00}(k_+,p_+,q_+)  
[1+n_T(\varepsilon_\mathbf{k})][1+n_T(\varepsilon_\mathbf{p})][1+n_T(\varepsilon_\mathbf{q})] \\
\mathcal{S}_{-++} &=& \mathcal{P}_T^{00|00|00}(k_-,p_+,q_+)       
 n_T(\varepsilon_\mathbf{k})[1+n_T(\varepsilon_\mathbf{p})][1+n_T(\varepsilon_\mathbf{q})]  \\
& \cdots \\
\mathcal{S}_{---} &=& \mathcal{P}_T^{00|00|00}(k_-,p_-,q_-) \, n_T(\varepsilon_\mathbf{k}) 
\, n_T(\varepsilon_\mathbf{p}) \, n_T(\varepsilon_\mathbf{q}) \, .
\end{eqnarray*}
where $k_\pm = \pm \varepsilon_{\bf k}$, and similarly for $p$ and $q$. We can then 
integrate in $\mathbf{p}$ and $\mathbf{q}$ to get:
\be\label{kt_integral}
 k_t(T) = -\frac{1}{(2\pi)^3} \int \di^3 {\rm k} \; \frac{1}{8\varepsilon_\mathbf{k}}  
  \frac{\partial^2}{\partial p_3 \, \partial q_3} \left[ \frac{1}{\varepsilon_\mathbf{p}\varepsilon_\mathbf{q}} 
\left( \mathcal{S}_{+++}+\cdots+\mathcal{S}_{---} \right) \right]_{\mathbf{p}=-\mathbf{k}, 
\, \mathbf{q}=\mathbf{k}} \, .
\ee

All the correlators in eq.~(\ref{correlat}) can be calculated in a similar fashion 
although it should be pointed out that the case of $k_t$ is somewhat simpler 
because in eq.~(\ref{kt_starting_point}) only one term in eq.~(\ref{JJT}) survived. 
Indeed, in general, one can have up to four terms associated with different sets 
of indices. Thus, the general correlator can be expressed as an integral of a 
sum of terms analogous to that appearing in eq.~(\ref{kt_integral}).

In the massless case, $T$ is the only energy scale and, on purely dimensional grounds, 
one finds that the correlators are of the form $\kappa(\xi)T^4$. For instance, 
integrating the eq.~(\ref{kt_integral}) with $m=0$ one obtains:
$$
k_t(T)=\left(-\frac{1}{30}\pi^2+\frac{1}{6}-\xi\right)T^4 \, .
$$
For the massive case, the integration is just a little more involved. First, the
angular part of the integration in ${\bf k}$ can be readily carried out and one
is left with expressions like:
\be\label{ikmt}
\frac{1}{2\pi^2} \int_0^{\infty} \!\!\mathrm{d} k \,\, I(k,m,T) \,.
\ee
where the function $I(k,m,T)$ is reported in table~\ref{integrand} for the various
correlators.
\begin{table*}
\caption{\label{integrand} The integrand functions $I(k,m,T)$ (see eq.~(\ref{ikmt}))
for the correlators in (\ref{correlat}) of a free real scalar field.}
\begin{ruledtabular}
\begin{tabular}{rl}
                 & $I(k,m,T)$ \\ \hline
$k_t$ & ${\frac{k^2}{96 T^2 \varepsilon_k}} 
\sinh^{-6}(\frac{\varepsilon_k}{2T}) \sinh(\frac{\varepsilon_k}{T}) \{k^2\varepsilon_k^2 + T^2 
[k^2 + 3 \varepsilon_k^2 (1 - 4 \xi)] [\cosh(\frac{\varepsilon_k}{T})-1] - 2 T k^2 \varepsilon_k 
\sinh(\frac{\varepsilon_k}{T}) \}$ \\
$k_\theta$ & $\frac{k^2}{288 T^2 \varepsilon_k} \sinh^{-6}(\frac{\varepsilon_k}{2T}) 
\sinh(\frac{\varepsilon_k}{T}) \{k^4 + 3 T^2 [k^2 (1 - 4 \xi) + \varepsilon_k^2 (8 \xi - 1)] 
[\cosh(\frac{\varepsilon_k}{T})-1] - 2 T k^2 \varepsilon_k \sinh(\frac{\varepsilon_k}{T}) \}$ \\
$k_s$ & $\frac{k^2}{480 T^2 \varepsilon_k} \sinh^{-6}(\frac{\varepsilon_k}{2T}) \sinh(\frac{\varepsilon_k}{T}) 
\{k^2\varepsilon_k^2 + 15 T^2 \varepsilon_k^2 (1 - 2 \xi) [\cosh(\frac{\varepsilon_k}{T})-1] 
- 5 T k^2 \varepsilon_k \sinh(\frac{\varepsilon_k}{T}) \}$ \\
$j_t$ & ${\frac{k^4}{24 \varepsilon_k}} \sinh^{-4}(\frac{\varepsilon_k}{2T}) \sinh(\frac{\varepsilon_k}{T}) (1 - 4 \xi)$ \\
$j_\theta$ & ${\frac{k^4}{72 \varepsilon_k}} \sinh^{-4}(\frac{\varepsilon_k}{2T}) \sinh(\frac{\varepsilon_k}{T}) (8 \xi - 1)$ \\
$j_s$ & ${\frac{k^4}{48 \varepsilon_k}} \sinh^{-4}(\frac{\varepsilon_k}{2T}) \sinh(\frac{\varepsilon_k}{T}) (2 \xi - 1)$ \\
$l_v$ & ${\frac{k^4}{24 T \varepsilon_k}} \sinh^{-3}(\frac{\varepsilon_k}{2T}) \cosh(\frac{\varepsilon_k}{2T}) 
[2 T (2 \xi - 1) + \varepsilon_k \coth(\frac{\varepsilon_k}{2T})]$ \\
\end{tabular}
\end{ruledtabular}
\end{table*}
The integral in eq.~(\ref{ikmt}) can be computed setting $k = m \sinh y$, which 
makes it possible to extract a $m^4$ factor; the integral then depends on $m$ and 
$T$ only through the ratio $x = m/T$ and one is then left with an adimensional integral 
over $y$ that can be turned into a series of type (\ref{thermal_function_series}) 
involving the modified Bessel functions of the second type $K_n(x)$:
\begin{equation}\label{thermal_function_series}
\frac{m^4}{2\pi^2}\sum_{r=1}^\infty a_r(x,\xi) \,,
\end{equation}
where, as has been mentioned, $x=m/T$. The final expression of functions $a$ and $\kappa$
can be found in table~\ref{table1}.
\begin{table*}
\caption{\label{table1}
The correlators (\ref{correlat}) calculated for a free real scalar field with vanishing
chemical potential. Also shown the well-known expressions of $\rho$ and $p$.}
\begin{ruledtabular}
\begin{tabular}{rrl}
& $\vphantom{\Big|}\kappa(\xi)$ & $a_r(x,\xi)$ \\ \hline
$\vphantom{\Big|}\rho$ & $\frac{1}{30}\pi^2$ & $-(rx)^{-2}K_2(rx) + (rx)^{-1}K_3(rx)$ \\
$\vphantom{\Big|}p$ & $\frac{1}{90}\pi^2$ & $(rx)^{-2}K_2(rx)$ \\
$\vphantom{\Big|}k_t$ & $-\frac{1}{30}\pi^2+\frac{1}{6}-\xi$ & $\frac{1}{12} 
\left\{[r^2-1+24\xi x^{-2}]K_2(rx) + 3[r(1-8\xi)-3r^{-1}]x^{-1}K_3(rx)\right\}$ \\
$\vphantom{\Big|}k_\theta$ & $-\frac{1}{90}\pi^2-\frac{1}{18}+\frac{1}{3}\xi$ & 
$\frac{1}{12} \left\{8(1-5\xi)x^{-2}K_2(rx)+[r(16\xi-3)-3r^{-1}]x^{-1}K_3(rx)\right\}$ \\
$\vphantom{\Big|}k_s$ & $-\frac{1}{90}\pi^2+\frac{1}{12}-\frac{1}{2}\xi$ & 
$-\frac{1}{4} \left\{ 2(1-2\xi)x^{-2}K_2(rx) + [r(4\xi-1)+r^{-1}]x^{-1}K_3(rx) \right\}$ \\
$\vphantom{\Big|}j_t$ & $\frac{1}{6}\left(1-4\xi\right)$ & $(1-4\xi) x^{-2}K_2(rx)$ \\
$\vphantom{\Big|}j_\theta$ & $\frac{1}{18}\left(8\xi-1\right)$ & $\frac{1}{3}(8ξ\xi-1) x^{-2}K_2(rx)$ \\
$\vphantom{\Big|}j_s$ & $\frac{1}{12}\left(2\xi-1\right)$ & $\frac{1}{2}(2\xi-1) x^{-2}K_2(rx)$ \\
$\vphantom{\Big|}l_v$ & $\frac{1}{135}\pi^2+\frac{1}{18}+\frac{1}{3}\xi$ & $\frac{1}{6} 
\left[(12\xi-6)x^{-2}K_2(rx) + (2r+r^{-1})x^{-1}K_3(rx)\right]$ \\
\end{tabular}
\end{ruledtabular}
\end{table*}

With the correlators calculated, we are now in a position to write down the coefficients 
of eq.~(\ref{udaw}), reported in table~\ref{table2} alongside with their non-relativistic 
limit $m/T = x \gg 1$, factorized as $n f(m,T)$ where:
\be\label{density}
n = \frac{m^3}{2\pi^2}\sum_{r=1}^\infty (rx)^{-1} K_2(rx)
\ee
is the particle density at the homogeneous equilibrium. The non-relativistic limit can 
be extracted by simply taking the asymptotic expansion of the $r=1$ term of each series. 

\begin{table*}
\caption{\label{table2}The coefficients of the stress-energy tensor in eq.~(\ref{texpa4}) 
calculated for a free real scalar field with vanishing chemical potential.}
\begin{ruledtabular}
\begin{tabular}{rrlr}
& $\vphantom{\Big|}\kappa(\xi)$ & $a_r(x,\xi)$ & $f(m,t)$ \\ \hline
$\vphantom{\Big|}U_\alpha$ & $\frac{1}{12}(1-6\xi)$ & $\frac{1}{24} 
\left[(r^2+24\xi x^{-2})K_2(rx) + 3(1-8\xi)rx^{-1}K_3(rx)\right]$ & $\frac{1}{24}m^2T^{-1} + 
\frac{1}{8}m(1-8\xi) + (\frac{5}{16}-\frac{3}{2}\xi)T + o(T)$ \\
$\vphantom{\Big|}U_w$ & $\frac{1}{12}(1-4\xi)$ & $\frac{1}{2}(1-4\xi)x^{-2}K_2(rx)$ & 
$(\frac{1}{2}-2\xi)T + o(T)$ \\
$\vphantom{\Big|}D_\alpha$ & $\frac{1}{18}(6\xi-1)$ & $\frac{1}{24} \left[(12-48\xi)x^{-2}K_2(rx) 
+ (24\xi-5)rx^{-1}K_3(rx)\right]$ & $m(\xi -\frac{5}{24})+(\frac{1}{2}\xi-\frac{1}{48})T + o(T)$ \\
$\vphantom{\Big|}D_w$ & $\frac{1}{6}\xi$ & $\xi x^{-2}K_2(rx)$ & $\xi T + o(T)$ \\
$\vphantom{\Big|}A$ & $\frac{1}{12}(1-6\xi)$ & $\frac{1}{4} \left[(4\xi-2)x^{-2}K_2(rx) + 
(1-4\xi)rx^{-1}K_3(rx) \right]$ & $m(\frac{1}{4}-\xi)+(\frac{1}{8}-\frac{3}{2}\xi)  T + o(T)$ \\
$\vphantom{\Big|}W$ & $\frac{1}{12}(2\xi-1)$ & $\frac{1}{2}(2\xi-1) x^{-2}K_2(rx)$ & 
$(\xi -\frac{1}{2}) T + o(T)$ \\
$\vphantom{\Big|}G$ & $\frac{1}{36}\left(1+6\xi\right)$ & 
$\frac{1}{6}\left[(6ξ\xi-3) x^{-2}K_2(rx) + rx^{-1}K_3(rx)\right]$ & $\frac{1}{6}m+(\xi -\frac{1}{12}) T+o(T)$ \\
\end{tabular}
\end{ruledtabular}
\end{table*}

As it can be seen from the table~\ref{table2}, all the coefficients $U, A, D, W, G$ 
have a finite non-relativistic limit with the dimension of an energy per unit volume. 
Consequently, as it has been mentioned, all the corrections to the stress-energy tensor 
in eq.~(\ref{texpa4}) are of quantum origin as they linearly depend on $\hbar$. 

The coefficient $W = \lambda_3/T^2$ for the massless case turns out to be in agreement 
with the calculation in ref.~\cite{moore} for $\xi=0$. However, unlike therein argued, 
we found that it has an explicit dependence on $\xi$, that is on the stress-energy 
tensor form.

\section{Thermodynamical inequivalence, frame dependence and equation of state}
\label{discuss}

We are now going to discuss some physical consequences of the general form of the 
stress-energy tensor (\ref{texpa4}) which we rewrite here:
\bea\label{texpa5}
 T^{\mu\nu}(x) = && \left[ \rho + \bar a^2 U_\alpha + \bar\omega^2 U_w \right] 
 u^\mu u^\nu -  
 \left[ p + \bar a^2  D_\alpha + \bar\omega^2 D_w \right] \Delta^{\mu\nu} \nonumber \\
 && + A \bar a^2 \hat a^\mu \hat a^\nu +  W \bar\omega^2 \hat\omega^\mu \hat\omega^\nu +  
  G \bar a \bar\omega (u^\mu \hat\gamma^\nu + \hat\gamma^\mu u^\nu) + o(\varpi^2)
\eea
where the shorthands $\bar a = \hbar |a|/cKT$ and $\bar\omega = \hbar |a|/KT$ for
the adimensional scales related to acceleration and vorticity. 

The first remarkable consequence is that, as pointed out in refs.~\cite{becatinti1,
becatinti2}, the mean stress-energy tensor in a general thermodynamic equilibrium
depends on the fundamental stress-energy tensor operator written in terms of the 
quantum fields. This is at variance with the familiar homogeneous equilibrium, and it
is made apparent by the dependence of the thermal functions other than $\rho$ and
$p$ in table \ref{table2} on the parameter $\xi$. If one was able to measure one 
of the coefficients multiplying $\bar a^2$ or $\bar \omega^2$ with a thermodynamics 
experiment, one would obtain information about the true, physical stress-energy 
tensor operator, hence on the correct gravitational theory, a conclusion already 
drawn in ref.~\cite{becatinti1}.

The second consequence is that, as it is apparent from the eq.~(\ref{texpa5}), $u^\nu = T 
\beta^\nu$ is not an eigenvector of $T^{\mu\nu}$ if $\gamma$ is non-vanishing, that 
is if the three vectors $\alpha,w,u$ (or $a,\omega,u$) are linearly independent, 
as it can be seen from the eq.~(\ref{texpa4}). This is what happens for the the 
rigid rotation, where $a$,$\omega$ and $u$ are orthogonal to each other. In this 
case, the $u$ vector does not coincide with the Landau definition of four-velocity, 
and should then be taken as defining a new hydrodynamical frame, dubbed the $\beta$
frame, as it has been extensively discussed in ref.~\cite{becalocal}.

The third, and perhaps the most striking consequence, is that the dependence of energy 
density and pressure on the temperature and chemical potential are modified with 
respect to the homogeneous equilibrium case. Also, there are more second-order coefficients 
in the expansion of the stress-energy tensor than previously envisaged. Looking at 
the eq.~(\ref{texpa5}) it can be realized that, with respect to the expansions presented 
in refs.~\cite{roma1,roma2,moore}, there are three new coefficients, that is $G, U_\alpha, 
U_w$ and two of them imply a modification of the energy density. One could argue 
that they would disappear by going to the Landau frame. Yet, in the diagonalization
of the stress-energy tensor in eq.~(\ref{texpa5}), it can be readily shown that,
retaining only quadratic terms in $\bar a$ and $\bar \omega$:
\bea\label{effective}
 \rho_{\rm eff} &=& \rho + \bar a^2 U_\alpha + \bar\omega^2 U_w + o(\varpi^2) \nonumber \\
  p_{\rm eff} &=& p + \bar a^2  \left( D_\alpha + \frac{1}{3} A \right)  
  + \bar\omega^2 \left( D_w + \frac{1}{3} W \right) + o(\varpi^2), 
\eea
where the effective pressure has been defined as the mean of the eigenvalues of 
the spacelike eigenvectors. Therefore, the energy density and the pressure coincide, in this
approximation, with those in the $\beta$ frame and the coefficients $U_\alpha$ 
and $U_w$ survive. One may wonder whether the modification of the energy density 
could be reabsorbed by a redefinition of the temperature other than the length 
of the $\beta$ vector in the density operator in the eq.~(\ref{gener1}), which is 
based on the maximization of entropy with macroscopic constraints \cite{becalocal}. 
In fact, a redefinition would cure only one of the eigenvalues of the stress-energy 
tensor, unless the coefficients $U,D,A,W$ fulfilled some preculiar relations. 
In all other cases, the relation between the eigenvalues of the stress-energy 
tensor, or the relation between energy density and pressure, in other words the 
equation of state $p_{\rm eff}(\rho_{\rm eff})$, is modified with respect to the 
homogeneous equilibrium case. For instance, in the non-relativistic limit of the 
massive case $m \gg T$ one has, according to table \ref{table2} that the leading 
corrections are those in $\bar a^2$, and restoring the natural constants:
\bea\label{effective2} 
 \rho_{\rm eff} &\simeq& \rho + \frac{1}{24} \frac{mc^2}{KT} \rho \bar a^2  
 = \left( 1 + \frac{1}{24} \frac{m \hbar^2 |a|^2}{(KT)^3} \right) \rho \nonumber \\
 p_{\rm eff} &\simeq& p + \left( \frac{2}{3}\xi - \frac{1}{8} \right) mc^2 \bar a^2 n = 
 p \left[ 1 + \left( \frac{2}{3}\xi - \frac{1}{8} \right) \frac{m \hbar^2 |a|^2}{(KT)^3} 
  \right]
\eea
where $\rho = m n$ and $p = n KT$ are the usual non-relativistic expressions for the
ideal Boltzmann gas and $n$ has the well known approximate expression:
$$
  n \simeq \left(\frac{m T}{2 \pi} \right)^{3/2} \e^{-m/T}
$$
We note in passing that the relations (\ref{effective2}) should hold in the case 
of a charged scalar field in the non-degenerate Boltzmann limit with a chemical
potential, that is:
$$
  n \simeq \left(\frac{m T}{2 \pi} \right)^{3/2} \e^{(\mu-m)/T}
$$
and negligible anti-particle contribution.

If it was possible to redefine $T$ to a new $T' = T + b(T) \bar a^2 $ such that 
$\rho = m n(T')$ and $p = T' n(T')$, then the coefficients in the $\bar a^2$ 
expansion of the functions would be the same. This can be shown by taking into 
account that $\partial n/\partial T \simeq (m/T^2) n(T)$ in the non-relativistic 
$m \gg T$ limit, so that
\begin{eqnarray*}
\rho(T') &=& m n(T') \simeq m n(T) + \frac{\partial n}{\partial T} (T'-T) =
 m n(T) \left( 1 + \frac{m}{T^2} b \bar a^2 \right) \nonumber \\
 p(T') &=& T' n(T') \simeq T n(T) + n(T) \left( 1 + \frac{m}{T} \right) b \bar a^2
  \simeq T n(T) \left( 1 + \frac{m}{T^2} b \bar a^2 \right)
\end{eqnarray*}
However, it can be seen by comparing the above equation with (\ref{effective2}) that
in general this is not the case, except when $\xi = 1/4$ which is neither the canonical 
nor the improved tensor. 

Furthermore, in general, the redefinition of a temperature would be mass dependent and it
would then be troublesome to define thermodynamic equilibrium at a common temperature
of a mixture of gases. Let
$$
\rho_{\rm eff}(T,\bar{a},\bar{\omega}) = \rho(T'(T,\bar{a},\bar{\omega})),
$$
where $\rho$ is the familiar homogeneous energy density. Expanding the new temperature
in $\bar a$ and $\bar\omega$ the leading order corrections must be of the second order:
$$
T'=T+T_{\bar{a}}(T)\bar{a}^2+T_{\bar{\omega}}(T) \bar{\omega}^2 + o(\varpi^2), 
$$
where $T_{\bar a}$ and $T_{\bar{\omega}}$ are proportional to the second derivatives 
of $T'(T,\bar{a},\bar{\omega})$ with respect to $\bar a$ and $\bar \omega$ respectively. 
These unknown functions can be obtained by comparing with the equation (\ref{effective}):
$$
\rho + \frac{\partial \rho}{\partial T}(T_{\bar{a}}\bar{a}^2+T_{\bar{\omega}} 
\bar{\omega}^2)=\rho + \bar a^2 U_\alpha + \bar\omega^2 U_w + o(\varpi^2),
$$
implying
$$
T'=T+\dfrac{U_\alpha}{\partial \rho/\partial T}\bar{a}^2 +\dfrac{U_w}{\partial 
\rho/\partial T}\bar{\omega}^2 + o(\varpi^2).
$$
Looking at the tables (\ref{table1}) and (\ref{table2}), it can be realized that 
the coefficients of $\bar a^2$ and $\bar\omega^2$ are non-trivial functions of the
mass and temperature.

Going now back to the properly defined $T = 1/\sqrt{\beta^2}$, we observe that, in 
the non-relativistic limit the relation between the effective energy density and 
pressure gets modified into:
$$
  p_{eff} \simeq \rho_{eff} \frac{KT}{m} \left[ 1 + \left( \frac{2}{3}\xi 
  - \frac{1}{6} \right) \frac{m \hbar^2 |a|^2}{(KT)^3} \right]
$$
Therefore, the effective equation of state depends on the acceleration besides 
the temperature. 
This could be surprising, but in fact in general global equilibrium all parameters, 
including acceleration and angular velocity play the role of thermodynamic 
variables on the same footing as temperature and chemical potential. It can be
seen that in the non-relativistic non-degenerate limit the quantum correction to 
the relations (\ref{effective}) and the equation of state becomes more important 
at low proper temperature, being proportional to $1/T^3$. Of course this applies
as long as the acceleration is such that $m \hbar^2 |a|^2/(KT)^3 \ll 1$ so that 
the expansion method holds \footnote{For a proton and $|a| = g$ one has that the 
ratio becomes ${\cal O}(1)$ for $T \approx 10^{-8}$ \{kelvin}; for very 
low temperatures, one would have to take more and more terms into account and 
eventually the exact solution would be needed.

\section{Conclusions}
\label{conclu}

In conclusion, we have demonstrated that the relativistic stress-energy tensor in 
general states of global thermodynamic equilibrium features quantum corrections 
with respect to its ideal form (\ref{tideal}) depending on the local values of acceleration 
and vorticity, besides proper temperature and chemical potential. We have calculated 
the coefficients of the additional terms of the stress-energy tensor in the appropriate
quantum statistical framework at the second order of an expansion in the parameters 
$\hbar a/cKT$ and $\hbar \omega/KT$ for the simplest case of a real scalar field. 
We have found that more terms exist with respect to previous assessments; our 
calculated coefficient $W$ for the real scalar field agrees with previous calculations 
\cite{moore}. 

We have emphasized three major physical consequences of this finding:
\begin{enumerate}
\item{} The coefficients explicitely depend on the form of the quantum stress-energy 
tensor operator, what was already argued in refs.~\cite{becatinti1,becatinti2}.
\item{} The effective energy density - defined as the eigenvalue of the stress-energy
tensor  - is also modified by terms involving acceleration and vorticity which
cannot be reabsorbed by means of a redefinition of the temperature.
\item{} The equation of state and the relation between effective pressure and effective
energy density are also modified by the presence of vorticity and acceleration.
\end{enumerate}

In principle, these findings could be extended to matter in local thermodynamic 
equilibrium in flat spacetime, as well as to matter in global/local equilibrium in 
a curved spacetime. In this case, it is well known that $\beta$ in eq.~(\ref{gener1}) 
must be a Killing vector which can have a non-vanishing exterior derivative $\partial_\mu 
\beta_\nu - \partial_\nu \beta_\mu$ and, consequently, additional terms of the
stress-energy tensor with respect to its ideal form (\ref{tideal}). This might be
of phenomenological relevance for the study of the equilibrium of self-gravitating 
objects.

\section*{Acknowledgments}

We are greatly indebted to R.~Panerai for numerous suggestions and help in 
calculations. We acknowledge interesting discussions with S.~Capozziello, N.~Pinamonti
and P.~Romatschke. 




\appendix

\section{$\wR$ expansion}

To derive the espressions of $\wR^{(n)}$, we can disregard, for the sake of simplicity, 
the conserved charge in (\ref{wrdef}) for it commutes with both $\wP$ and $\wJ$ 
operators. Defining
\be\label{abdef}
 \wA = - \beta_\mu  \wP^\mu \qquad \qquad 
 \wB = \frac{1}{2} \varpi_{\mu\nu} \wJ^{\mu\nu} \,,
\ee
and applying the known Poincar\'e algebra relations, we find:
\begin{align*}
[\wA, \wB] &= - i \beta^\mu  \varpi_{\mu\nu} \wP^\nu \,, \\
[[\wA, \wB], \wA] &= 0 \,, \\
[[\wA, \wB], \wB] &= \beta^\mu  \varpi_{\mu\nu} \omega_{\rho\sigma} g^{\nu\rho} 
 \wP^\sigma \,, \\
[[[[\wA, \wB], \wB], ...], \wB] &= -(i)^n \beta^\mu  (\varpi\cdot\varpi
 \cdot\ldots\cdot\varpi)_{\mu\nu} \wP^\nu \,.
\end{align*}
Now, using the Baker--Campbell--Hausdorff formula to expand $\exp[\wA+\wB]$ and 
retaining only the non-vanishing terms; taking into account that any commutator involving 
$\wA$, $\wB$ or commutators thereof, will in turn commute with $\wA$, being proportional 
to four-momentum operators, we obtain:
\bea\label{BCH}
 \wR(\beta,\varpi) = \e^{\wA+\wB} &\simeq& 
  \e^{\wB} \, \e^{\wA} \, \e^{\frac{1}{2!}[\wA,\wB]} \, \e^{\frac{1}{3!}[[\wA,\wB],\wB]} 
  \, \e^{\frac{1}{4!}[[[\wA,\wB],\wB],\wB]} \dots \nonumber \\
  &=& \e^{\wB} \, \e^{\wA+\frac{1}{2!}[\wA,\wB]+\frac{1}{3!}[[\wA,\wB],\wB]+
  \frac{1}{4!}[[[\wA,\wB],\wB],\wB]+\dots }
\eea
and its expansion up to second order in $\wB$ (which is tantamount to a second order 
in $\varpi$) reads:
\be\label{form1}
\wR(\beta,\varpi) \simeq \e^{\wA} + ( \wB + \frac{1}{2} [\wA,\wB] ) \e^{\wA} 
+ \left( \frac{1}{2}\wB^2 + \frac{1}{3}\wB [\wA,\wB] + \frac{1}{6}[\wA,\wB]\wB + 
 \frac{1}{8}[\wA,\wB]^2 \right) \e^{\wA} 
\ee
where advantage has been taken of the fact that $\exp[\wA]$ commutes with both the 
commutators $[\wA,\wB]$ and $[[\wA,\wB],\wB]$. Now, by using the relation:
$$
 \e^{-\wA} \wB \e^\wA = \wB - [\wA,\wB]
$$
which is a known corollary of the Baker--Campbell--Hausdorff formula for our case, 
the eq.~(\ref{form1}) can be rewritten as:
\be\label{form2}
 \wR(\beta,\varpi) \simeq \e^{\wA} + \e^{\wA} ( \wB - \frac{1}{2} [\wA,\wB] ) 
 + \e^{\wA} \left( \frac{1}{2}\wB^2 - \frac{1}{6} \wB [\wA,\wB] - 
 \frac{1}{3}[\wA,\wB ]\wB + \frac{1}{8}[\wA ,\wB ]^2  \right) 
\ee
We can now take the half-sum of (\ref{form1}) and (\ref{form2}) to obtain:
\be\label{formf}
 \wR(\beta,\varpi) \simeq \e^{\wA} + \frac{1}{2} \left\{ \e^{\wA}, \wB \right\} 
 + \frac{1}{4} \left\{ \e^{\wA},\wB^2 \right\} - \frac{1}{8} \e^{\wA}[\wA,\wB]^2
 - \frac{1}{12} \e^{\wA}[[\wA,\wB],\wB]
\ee
putting the expressions of $\wA$ and $\wB$ in eq.~(\ref{abdef}) in the eq.~(\ref{formf})
one can read off the operators in eq.~(\ref{expand}), which are quoted in 
eq.~(\ref{expand2}).

\section{Calculation of angular momentum-stress energy tensor correlators}

The density operator (\ref{homo}), which is used to calculate the mean values denoted 
as $\langle \; \rangle_{\beta}$ can be written as $\widehat{\sf \Lambda} \rho_0 
\widehat{\sf \Lambda}^{-1}$ where ${\widehat{\sf \Lambda}}$ is the Lorentz transformation
turning $\beta_0 = (1/T,{\bf 0})$ into $\beta$. Hence, the mean value of a general 
tensor can be expanded as: 
\be\label{reduce1}
  \langle \widehat O^{\mu_1,\ldots,\mu_N} \rangle_{\beta(x)} = 
  \Lambda^{\mu_1}_{\nu_1} \ldots \Lambda^{\mu_N}_{\nu_N} 
  \langle \widehat O^{\nu_1,\ldots,\nu_N} \rangle_T
\ee
where $\langle \; \rangle_T$, as has been mentioned in the text, stands for the 
mean value with the density operator $\frac{1}{Z} \exp[-\beta_0 \cdot \wP] = 
\frac{1}{Z}\exp[-\widehat H/T]$. Note that: 
\be\label{reduce2}
  \Lambda^\mu_0 = \hat \beta^\mu  \qquad \qquad 
  \sum_{i,j=1}^3 \Lambda^{\mu}_i \Lambda^{\nu}_j g^{ij} = 
  g^{\mu\nu} - \hat \beta^\mu \hat \beta^\nu = \Delta^{\mu\nu}
\ee
Since $\exp[-\widehat H/T]$ is invariant by rotation, only scalars under spatial 
rotation, either components or contractions of the tensor $O^{\nu_1,\ldots,\nu_N}$ may 
have a non vanishing value. Furthermore, we assume that the hamiltonian operator 
is symmetric under parity and time reversal transformations, so that also pseudoscalars
and scalars which are odd under time reversal will have vanishing mean value. 

For instance, for a symmetric tensor operator $\widehat S^{\mu\nu}$ one can write:
\be\label{symmt}
\langle \widehat S^{\mu\nu} \rangle_T = \delta^\mu_0 \delta^\nu_0 A + \Deltaz^{\mu\nu} B
\ee
where $\Deltaz^{\mu\nu}$ is the transverse projector in the rest frame, i.e.
$\Deltaz^{\mu\nu} = g^{\mu\nu} - \delta^\mu_0 \delta^\nu_0$. Of course, the 
eq.~(\ref{symmt}) becomes the well known:
$$
 \langle \widehat S^{\mu\nu} \rangle_{\beta} = A \hat\beta^\mu \hat\beta^\nu
 + \Delta^{\mu\nu} B 
$$
by using the (\ref{reduce1}) and (\ref{reduce2}). The coefficients $A$ and $B$
can be calculated from the mean values selecting the components in eq.~(\ref{symmt}) 
which make all terms vanishing except one. Thereby:
$$
  A = \langle \widehat S^{00} \rangle_T  \qquad\qquad 
  B = - \langle \widehat S^{ii} \rangle_T
$$

This general procedure can be applied to the calculations of tensors of any rank.
Indeed, in view of eq.~(\ref{reduce2}), anytime a time component or $0$ index is selected 
in $\langle \widehat O^{\nu_1,\ldots,\nu_N} \rangle_T$ in eq.~(\ref{reduce1}) a 
$\delta^\nu_0$ will appear eventually turning into a $u$ after boosting, while for 
a space contraction of indices a $\Deltaz$ projector will, eventually turning into 
a $\Delta$ like in eq.~(\ref{reduce2}).
 
We can first apply the above argument to the calculation of $\langle \wJ^{\mu\nu} 
\wT^{\rho\sigma} \rangle_{\beta}$. By using the decomposition (\ref{decomp2}) and 
taking into account (\ref{reduce1}), the only possible non-vanishing contributions 
read:
$$ 
{\rm Re} \langle \wK_i \wT^{0i} \rangle_{T} \qquad {\rm Re}\langle \wJ_i \wT^{0i} \rangle_{T}
$$
Yet, they both vanish because they are odd under time reversal and parity respectively. 
No scalar can be formed with $\langle \wJ^{\mu\nu} \rangle_T$ and so the mean value 
of the angular momentum $\langle \wJ^{\mu\nu} \rangle_\beta$ vanishes too.

Let us now move to the more complicated case of correlators involving two angular
momentum operators, starting from:
$$
  \langle \{\wK^\rho, \wK^\sigma \}; \wT^{\mu\nu} \rangle_T
$$
In the rest frame, $\{\wK^\rho,\wK^\sigma \}$ is a symmetric tensor with vanishing time 
components, so it has one spin-0 component obtained with the contraction of the indices 
$\rho$ and $\sigma$ and one spin-2 component under rotation which can be obtained by 
applying the projector:
$$
  P^{\rho\sigma}_{\alpha\beta} = \frac{1}{2} \left( 
  \Deltaz^{\rho}_{\alpha} \Deltaz^{\sigma}_{\beta} + \Deltaz^{\sigma}_{\alpha} 
  \Deltaz^{\rho}_{\beta} - \frac{2}{3} \Deltaz^{\rho\sigma} \Deltaz_{\alpha\beta} 
  \right)
$$  
to the tensor itself. In order to construct a rotation singlet, we need to combine 
the above components with the corresponding components of $\wT^{\mu\nu}$. The spin 0 
components can only contract with $\wT^{00}$ and its spatial trace, so one obtains
two contributions:
$$
-\Deltaz^{\rho\sigma} \delta^\mu_0 \delta^\nu_0 k_t 
\qquad \Deltaz^{\rho\sigma} \Deltaz^{\mu\nu} k_\theta 
$$
whereas the contraction of the spin 2 component of $\{\wK^\rho, \wK^\sigma \}$ 
with the one of $\wT$ gives rise to:
$$
\left( \Deltaz^{\rho\mu} \Deltaz^{\sigma\nu} + \Deltaz^{\rho\nu}\Deltaz^{\sigma\mu} 
-\frac{2}{3} \Deltaz^{\mu\nu}\Deltaz^{\rho\sigma} \right) k_s 
$$
Altogether
\be\label{keikei}
\frac{1}{2} {\rm Re} \langle \{\wK^\rho, \wK^\sigma \} ; \wT^{\mu\nu}\rangle_T 
 = -\Deltaz^{\rho\sigma} \delta^\mu_0 \delta^\nu_0 k_t + 
 \Deltaz^{\rho\sigma} \Deltaz^{\mu\nu} k_\theta + 
\left( \Deltaz^{\rho\mu}\Deltaz^{\sigma\nu} + \Deltaz^{\rho\nu}\Deltaz^{\sigma\mu} 
-\frac{2}{3} \Deltaz^{\mu\nu}\Deltaz^{\rho\sigma} \right) k_s 
\ee
which in the observer frame reads:
$$
\frac{1}{2} {\rm Re} \langle \{\wK^\rho, \wK^\sigma \} ; \wT^{\mu\nu}\rangle_\beta
 = -\Delta^{\rho\sigma} u^\mu u^\nu k_t + \Delta^{\rho\sigma} \Delta^{\mu\nu} k_\theta + 
\left( \Delta^{\rho\mu}\Delta^{\sigma\nu} + \Delta^{\rho\nu}\Delta^{\sigma\mu} 
-\frac{2}{3} \Delta^{\mu\nu}\Delta^{\rho\sigma} \right) k_s 
$$
To find a compact expression of the coefficients $k_t,k_\theta,k_s$ one can select
the indices making all terms on the right hand side of (\ref{keikei}) vanishing 
except the one of interest. One can check that all indices in the definitions 
(\ref{correlat}) are properly chosen (notice how in $k_s$ we avoided the symmetrization 
in $\mu\leftrightarrow\nu$ associated with the anti-commutator since we know that the 
anti-symmetric part will not contribute). For the $j_t,j_\theta,j_s$ the procedure 
is precisely the same outlined above with the replacement $\wK \to \wJ$. 

In fact, the correlator $\langle \{\wK^\rho,\wJ^\sigma\} ; \wT^{\mu\nu} \rangle_T$ 
is a somewhat special case because $\{\wK^\rho,\wJ^\sigma\}$ is odd under parity 
and time reversal. Therefore, the only non-vanishing contraction is between the two
spin-1 components of the tensors $\{\wK^\rho,\wJ^\sigma\}$ and $\wT^{\mu\nu}$ respectively.
The spin-1 components can be obtained by means of the projectors
$$
 P^{\rho\sigma}_{\alpha\beta} = - \frac{1}{2}
 \epsilon^{0\rho\sigma\tau}\epsilon_{0\alpha\beta\tau}
$$
and
$$
 P^{\mu\nu}_{\alpha\beta} = \frac{1}{2} \delta^\mu_0 \Deltaz^{\nu}_{\alpha}
  \delta^0_\beta + \delta^\nu_0 \Deltaz^{\mu}_{\alpha} \delta^0_\beta
$$
respectively. Hence:
$$
{\rm Re} \langle \{\wK^\rho,\wJ^\sigma\}; \wT^{\mu\nu} \rangle_T = 
 \left( \epsilon^{0\rho\sigma\tau} \delta^\mu_0 \Deltaz^{\nu}_{\tau} +
 \epsilon^{0\rho\sigma\tau} \delta^\mu_0 \Deltaz^{\nu}_{\tau} \right) l_v
 = \left( \delta^0_\kappa \epsilon^{\kappa\rho\sigma\nu} \delta^\mu_0 +
 \delta^0_\kappa \epsilon^{\kappa\rho\sigma\mu} \delta^\nu_0 \right) l_v
$$
which, once boosted, reads:
$$
{\rm Re} \langle \{\wK^\rho,\wJ^\sigma\}; \wT^{\mu\nu} \rangle_\beta 
 = \left( u_\kappa \epsilon^{\kappa\rho\sigma\nu} u^\mu +
 u_\kappa \epsilon^{\kappa\rho\sigma\mu} u^\nu \right) l_v
$$

\end{document}